\documentclass[twocolumn,showpacs,preprintnumbers,amsmath,amssymb]{revtex4-1}

\usepackage{graphicx}
\usepackage{amssymb}
\usepackage{hyperref}

\begin{document}

\title{First observation of Ce volume collapse in CeN$^*$}

\author{J.G. Sereni}
\address{Low Temperature Division CAB-CNEA, CONICET, 8400 S.C. de
Bariloche, Argentina}

\begin{abstract}

{On the occasion of the 80th anniversary of the first observation
of Ce volume collapse in CeN a remembrance of the implications of
that transcendent event is presented, along with a review of the
knowledge of Ce physical properties available at that time.
Coincident anniversary corresponds to the first proposal for Ce as
a mix valence element, motivating to briefly review how the
valence instability of Ce was investigated since that time.

$^*$ This review was presented in a plenary session at the SCES
2017 conference, Prague, 17 July 2017.}

\end{abstract}
\date{\today}

\maketitle


\section{\lowercase{2017 - 80th anniversary of the first observation
of }C\lowercase{e volume collapse in} C\lowercase{e}N}

\begin{figure}[tb]
\begin{center}
\includegraphics[width=19pc]{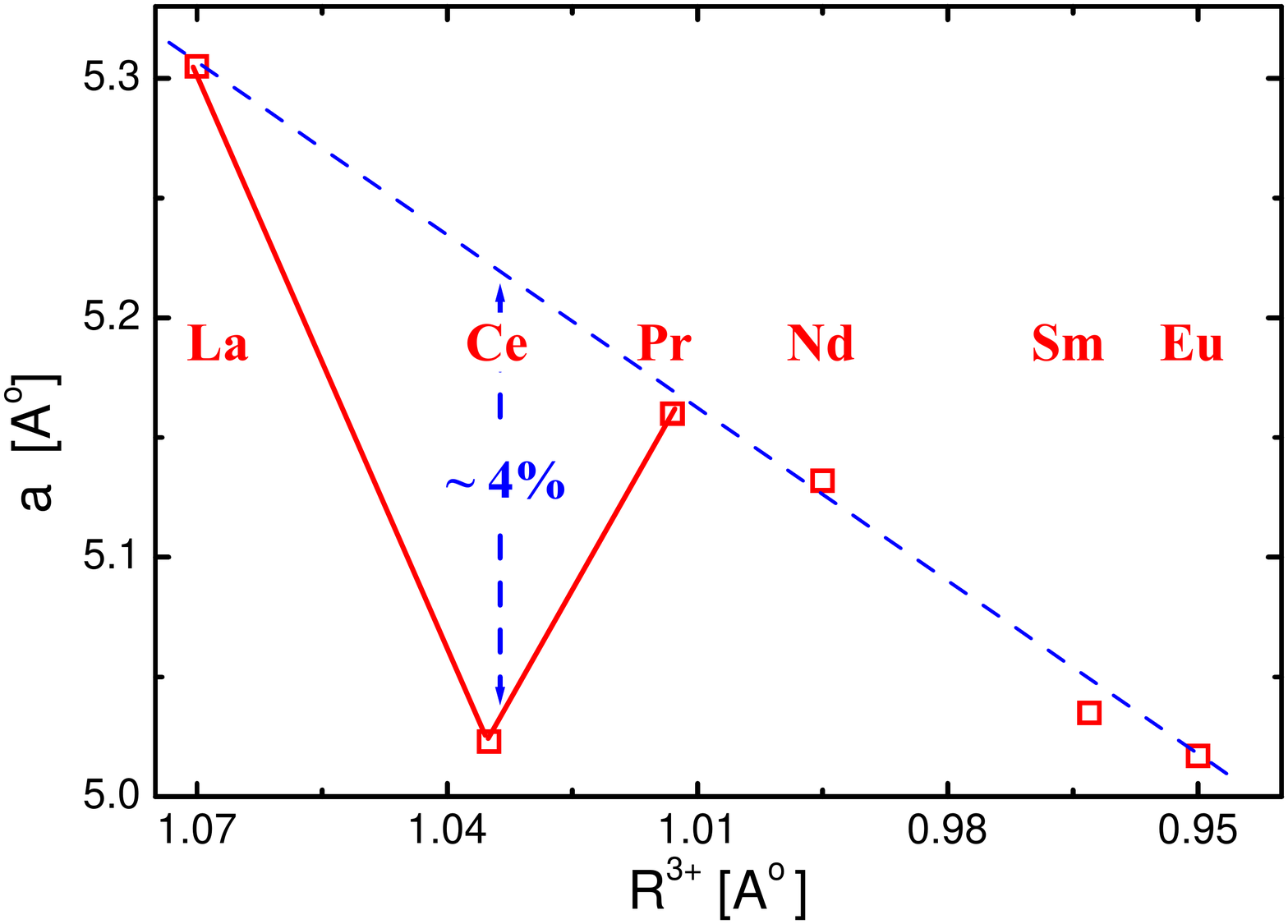}
\end{center}
\caption{Comparison of the lattice parameter of the rare earth
nitrides series as a function of the respective ionic radius.
Dashed line indicates the trend of the REN volume decrease. Data
for NdN, SmN and EuN after \cite{Natali}.} \label{F1}
\end{figure}

In Volume XXV, p.129 (1937), of the "Rendiconti Accademia
Nazionale dei Lincei", a report on the crystal structure of rare
earth nitrides (REN, RE = La, Ce, Pr and Nd) was presented. There,
the authors, Aldo Inadelli and Erico Botti of the Institute of
Chemistry of the University of Genova (Italy), informed that {\it
"The lattice parameter of CeN presents a strange anomaly"} in
comparison to the linear decrease of the other REN compounds. As
it can be seen in Fig.~\ref{F1}, the lattice parameter $`a'$ of
the NaCl-type structure of CeN ($a = 5.011\,\AA$) {\it "... is
much lower than that of the neighbor compound with Pr ($a =
5.155\,\AA$) and also to that of }[pure] {\it face centered cubic
Ce ($a = 5.141$\,\AA)"}. This reduction of the lattice parameter
in CeN is about 4\% of the value extracted from the interpolation
between the respective LaN ($a = 5.275\,\AA$) and PrN lattice
parameters, as depicted in Fig.~\ref{F1}.

To verify such unexpected collapse of the CeN volume, the authors
mention that {\it "For security, many photograms were performed on
different CeN samples, trying to change as much as possible the}
[compound]{\it formation conditions, always with the same
result."} They recognize that {\it "The abnormal behavior of Ce is
difficult to explain because the CeN compound does not present
different characteristic than other compounds"}. However, they
remark that {\it the only ascertained difference} [with the
neighboring LaN] {\it is that the heat of formation seems to be
larger."}

\begin{figure}[tb]
\begin{center}
\includegraphics[width=19pc]{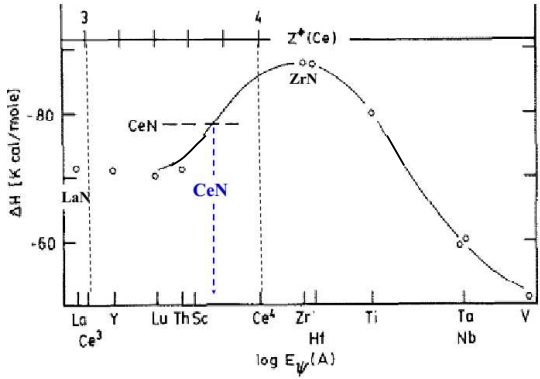}
\end{center}
\caption{Heat of formation of selected XN compounds as a function
of the logarithm of respective Electrostatic Energies of
X-elements, after \cite{JLCM82}.} \label{F2}
\end{figure}

This observation is backed by the fact that a large volume
contraction ($\Delta V_f$) respect to the volume of elementary
components in the compound formation process corresponds to a
large heat of formation ($\Delta H_{f}$) because $\Delta H_f
\propto \Delta V_f$ \cite{DHDV}. This feature is illustrated in
Fig.~\ref{F2} \cite{JLCM82} where $\Delta H_{f}$ is plot as a
function of the Electrostatic Energy $E_{\phi} = q \phi$
\cite{tarcho}, being $\phi \propto q Z/r * exp(-\lambda/r)$ the
electrostatic potential including the Thomas-Fermi screening
factor: $exp(-\lambda/r)$. The $\Delta H_{f}$ values for
La$^{3+}$N, CeN and Zr$^{4+}$N are taken from Ref. \cite{Miedema}.
From Fig.~\ref{F2} it becomes clear that the extra $\Delta H_{f}$
observed in CeN is related to the shift of the Ce$^{3+}$ valence
$\nu = 3$ in the pure metal towards $\nu = 4$, reaching a value of
$\nu \approx 3.7$.

Nowadays, the extraordinary volume contraction or large $\Delta
H_{f}$ of CeN compared with LaN and PrN can be understood taking
into account the transference (or delocalization) of the $4f^1$-Ce
electron to the electronic band, which is related to the drastic
change of valence.

\section{W\lowercase{hat was known before 1937 concerning }C\lowercase{e and
}C\lowercase{e-alloys properties}}

In its elemental form, Ce was first isolated by Carl G. Mosander
at the Chemical Laboratory of the Caroline Medical Institute,
Stockholm, in {\bf 1839} \cite{Mosander} while identifying the
mayor components of mischmetal alloys \cite{Mosand2}.

An early report on Ce alloys with transition elements (Fe, Co and
Ni) appeared {\bf 1903} by Auerton Welsbach \cite{1903} .
According the author, {\it "the property of giving off sharp
sparks with a hard object on the basis of the iron alloys of
cerium is also peculiar to its alloys with cobalt, nickel and
manganese"}. Shortly after, {\bf 1904}, W. Muthmann and H. Beck
\cite{1904} published their studies on the heat of formation of
Ce-Mg alloys. These thermochemical studies are followed by those
performed on CeN by J. Lipzki in {\bf 1909} \cite{1909}.

\begin{figure}[tb]
\begin{center}
\includegraphics[width=19pc]{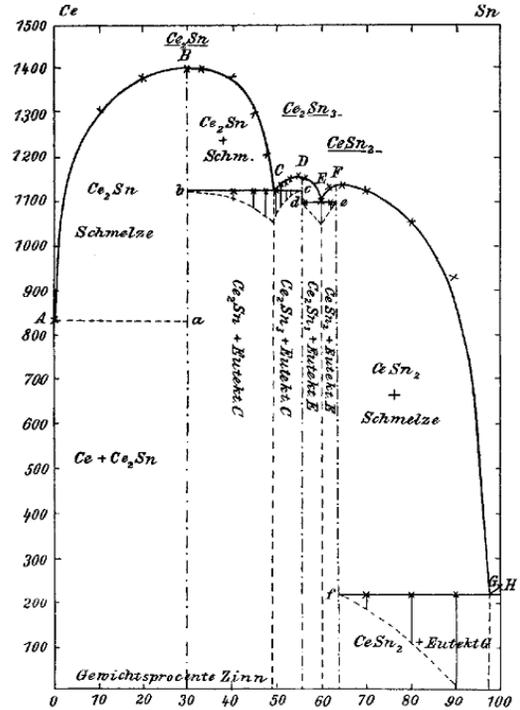}
\end{center}
\caption{Phase diagram of the Ce-Sn binary alloys published in
1911 \cite{1911}.} \label{F3}
\end{figure}

Two extensive articles on Ce-Sn and Ce-Al alloys containing the
early phase diagrams of these binary systems were published by
Rudolf Vogel, from the Chemical Physics Institute of the
University of G\"otingen, in {\bf 1911} \cite{1911}, see
Fig.~\ref{F3} and {\bf 1912} \cite{1912} respectively. These
papers contain a detailed description of the samples preparation
procedure.

\subsection{Cerium properties as pure metal, at ambient conditions}

\begin{figure}[tb]
\begin{center}
\includegraphics[width=19pc]{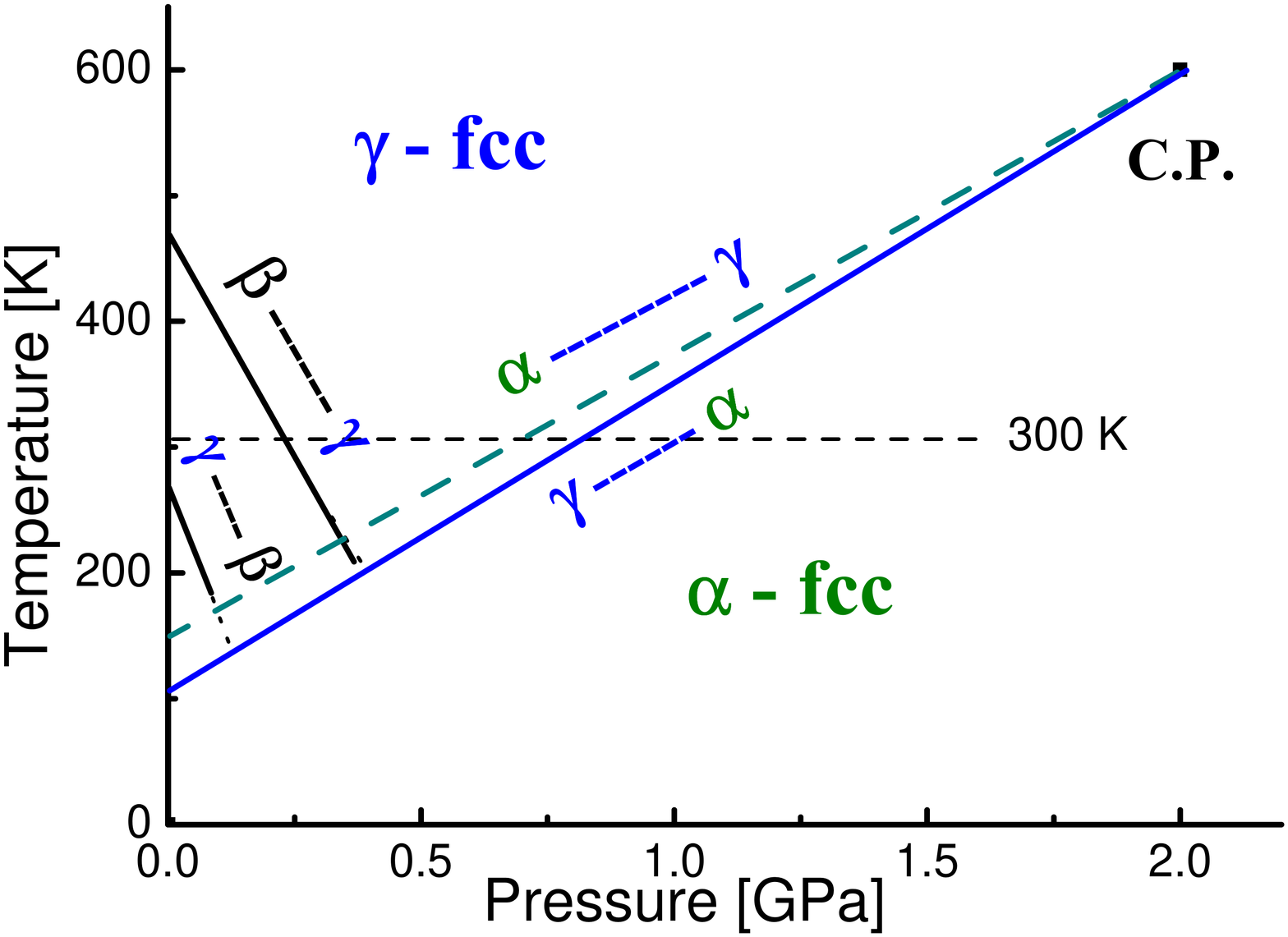}
\end{center}
\caption{(Color online) Phase diagram of pure Cerium under
hydrostatic pressure, after \cite{Gschneidner62}. Dashed line
indicates ambient temperature.} \label{F4}
\end{figure}

In the {\bf crystal structure} study performed by Albert W. Hull,
from General Electric Res.Lab. in {\bf 1921} \cite{Hull}, he
reports that {\it "Cerium shows the same structure as Ti and Zr
[hex] with axial ratio 1.62"} and {\it "The side of the elementary
triangle is $3.65\,\AA$ and height $5.96\,\AA$."}. However, he
notices that {\it "There is present also a face centered cubic
[$\gamma$ - fcc] form, with side of cube $5.12\,\AA$ giving the
same density as the hexagonal form. Cerium thus appears to be
composed of a mixture of the hexagonal and cubic forms of
close-packed."}, see Ce metal phase diagram in Fig.~\ref{F4} at
ambient pressure.

There is a work on {\bf elastic properties} of heavy elements by
P. W. Bridgman, reported in {\bf 1927} with the title "The
compressibility and pressure coefficient of ten elements"
\cite{Brigdman1927}. This is the first observation of the $\gamma
- \alpha$ transition at room temperature performed through
electroresistence measurements, see the 300\,K isotherm in
Fig.~\ref{F4}.

\begin{figure}[tb]
\begin{center}
\includegraphics[width=19pc]{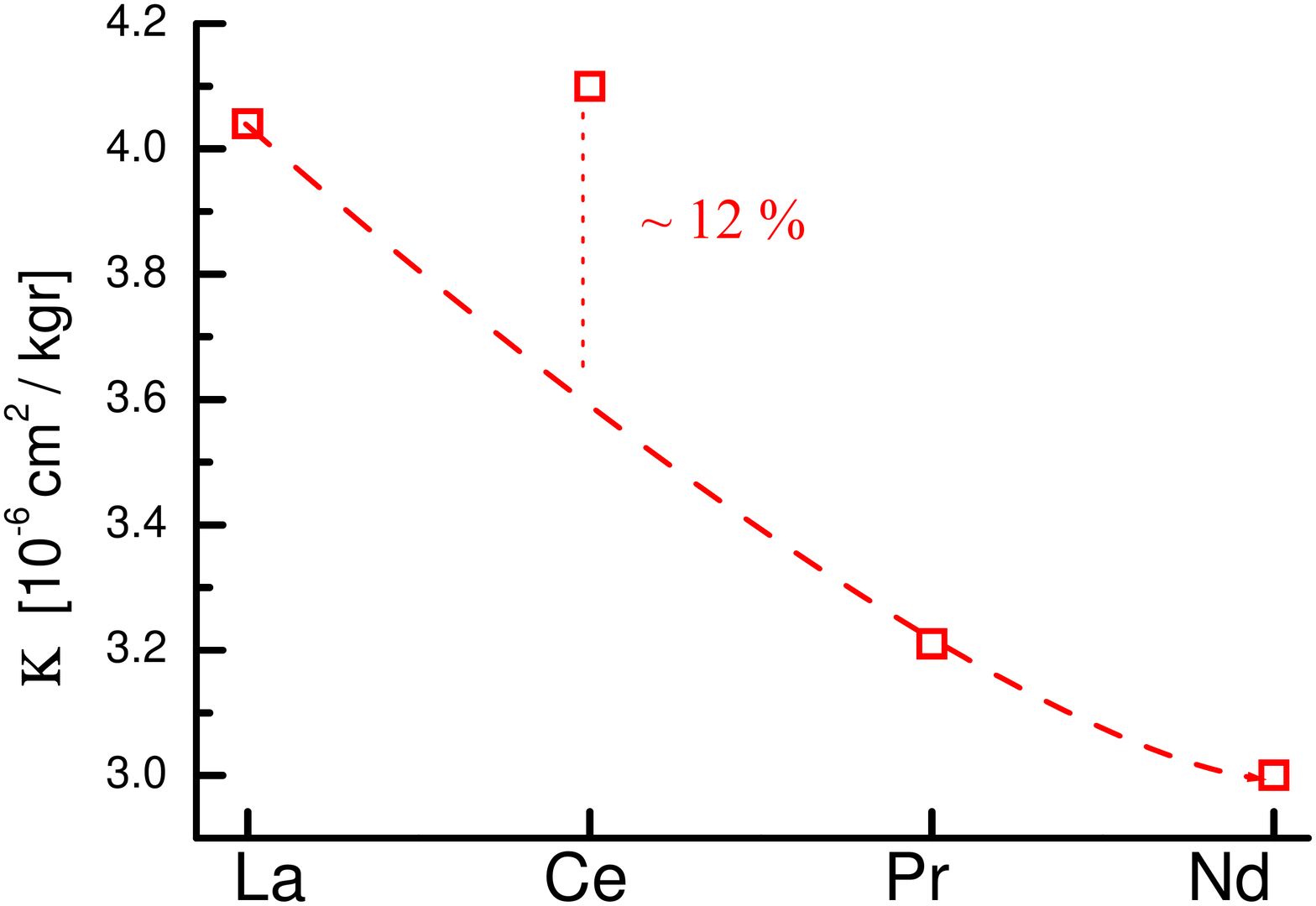}
\end{center}
\caption{(Color online) Comparison of pure Ce compressibility with
the neighboring rare earth elements, data from Ref.\cite{CRC}.}
\label{F5}
\end{figure}

In {\bf 1932} the existence of a cubic allotropic Ce phase was
confirmed by Von Laurence L. Quill, at the University of
G\"ottingen (Germany) \cite{Quill1932}. The author confirms that
{\it "Ce still continuous crystallizing, according to A.W. Hill,
in face-centered-cubic"}, and remarks that
{\it "Ce shows a higher compressibility than La and Pr." }, see
Fig.~\ref{F5}. The Ce lattice parameters reported in its different
allotropic phases are: i) for the hexagonal structure
$a=3.65\,\AA$ and $c=5.96\,\AA$, with a ratio $c/a=1.633$ and ii)
for the cubic structure $a=5.12\,\AA$.

\subsection{Ce properties as a function of temperature}

\begin{figure}[tb]
\begin{center}
\includegraphics[width=19pc]{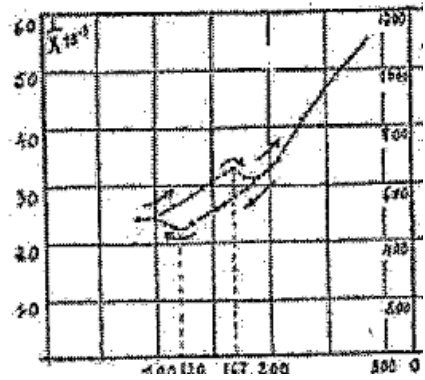}
\end{center}
\caption{Inverse susceptibility of Ce metal showing an hysteretic
trajectory between 100 and 200\,K.} \label{F6}
\end{figure}

The first study on the {\bf magnetic properties} of pure Ce were
published in {\bf 1934} by M.F. Trombe to the French Academy of
Sciences in a work presented by M.P. Weiss under the title {\it
"Propri\'et\'es magn\'etiques du Ce, du La et du Nd \`a diverses
temperatures"} \cite{Trombe1934}. The author reports that {\it "At
low temperature, Ce possesses complex magnetic properties: i) at
99\,K its magnetization coefficient drops from $53.8 \time
10^{-6}$ (under $H = 3$\,kGs) to $ 39 \time 10^{-6}$ (under $H =
7$\,kGs). ii) It seems to have two states, with different
temperatures for heating and cooling}, see Fig.~\ref{F6}.

One decade later {\bf 1944} a new research on the {\it "Range of
existence and properties of the allotropic states of metallic
cerium"} studied through {\bf thermal expansion} measurements was
published by F. Trombe and M. Foex \cite{Trombe44}. They mention
that the Ce-$\alpha$ state, which has previously been defined by
its thermomagnetic anomaly was studied by dilatometry and that
{\it "It can be shown that the magnetic variation almost coincides
with an exceptionally great anomaly of expansion which amounts to
10\% in volume."}. Their measurements show that {\it ".. during
contraction (decreasing temperature) the expansion coefficient has
a maximum at -164\,$^o$C (109.3\,K) whereas, with increasing
temperature, the maximum is reached at -98\,$^o$C (175.4\,K). The
maximums of the magnetic anomaly have been found at  -163\,$^o$C
(108.4\,K) and -100\,$^o$C (177.4\,K) respectively"}. These
results support the Ce-magnetic phase diagram presented in
Fig.~\ref{F4}.

\begin{figure}[tb]
\begin{center}
\includegraphics[width=21pc]{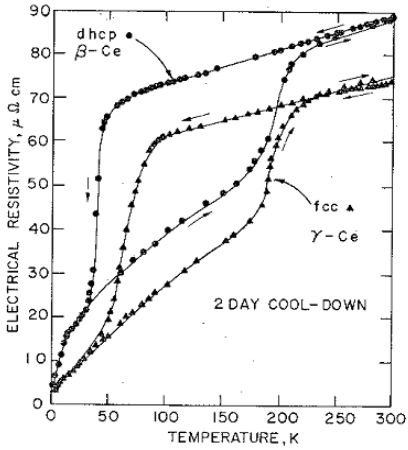}
\end{center}
\caption{(Color online) Electrical resistivity of pure Ce as a
function of temperature showing different hysteresis cycles when
starting from stabilized dhcp-$\beta$ or fcc-$\gamma$ phases,
after \cite{Gschneidner76}}. \label{F7}
\end{figure}

The mentioned magnetic and dilatometric behavior were then
confirmed by {\bf electrical conductivity} measurements, which
indicate that {\it "The electrical conductivity tests were made
with a 99.6\% pure Ce, ....,  cooled to low temperature, where it
passes into the $\alpha$ state, shows an electrical anomaly at
-186° (87.4\,K) and -98° (175.4\,K), respectively}. Fig.~\ref{F7}
\cite{Gschneidner76} represents those results

\begin{figure}[tb]
\begin{center}
\includegraphics[width=20pc]{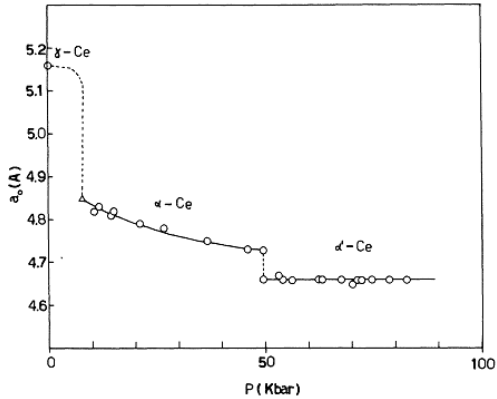}
\end{center}
\caption{Lattice parameters of different allotropic Ce metallic
phases: $a_0$\,[\AA], as a function of pressure [kbar] after
\cite{Franceschi69}.} \label{F8}
\end{figure}

New results on {\bf pressure effects} investigated by P.W.
Bridgman in {\bf 1948} were quoted by A.W. Lawson \cite{Lawson49}.
Those results indicate an over-all contraction in volume of 16.5\%
at 15,000 atmospheres ($\approx 15$\,kbar) \cite{Bridgman48}. This
contraction is illustrated in Fig.~\ref{F8} as the evolution of
Ce-lattice parameter with pressure.

\section{Valence Instabilities}

\section{\lowercase{80th anniversary of the report on Valence Instabilities in
Lanthanid elements}}

\begin{figure}[tb]
\begin{center}
\includegraphics[width=21pc]{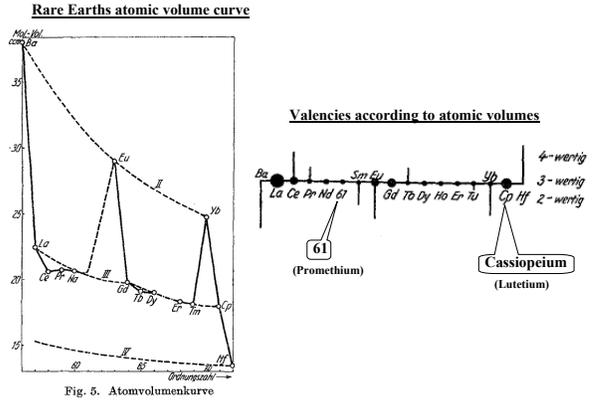}
\end{center}
\caption{(Color online) (left) Volume contraction of the
lanthanide series as a function of the atomic number. (right)
Lanthanides valencies at ambient conditions, after
\cite{Klemm37}.} \label{F9}
\end{figure}

Another relevant anniversary corresponds to the review article by
Von W. Klemm and H. Bommer in {\bf 1937} \cite{Klemm37}. This work
was addressed to the volume contraction of the lanthanide series
as a function of the atomic number, Fig.~\ref{F9}-left, and their
different electronic configurations at ambient conditions,
Fig.~\ref{F9}-right. As it can be seen in Fig.~\ref{F9}-left Eu
and Yb strongly expand their atomic volumes by reaching their
divalent (2$^+$) configurations. On the contrary, Ce atoms
slightly reduces its volume respect to the other lanthanides as it
increases its valence. On the right side of Fig.~\ref{F9} a
collection of extreme deviations from the trivalent (3$^+$) states
of the RE element is presented where also Sm is included into the
2$^+$ tendency whereas Pr and Tb into the 4$^+$ direction.
Noteworthy is the name of 'Cassiopeium' for present Lu and the
number '61' to identify Promethium not yet isolated at that time.

\subsection{Ce as a 'static mix' of 3$^+$/4$^+$ electronic configurations}

In chapter 3 of Ref.\cite{Klemm37} under the title "Rare Earth
metals ionic moments" the authors state the {\it "For Ce, from the
atomic volumetric curve we conclude that there are both three and
tetra valent ions. This assumption is magnetically easy to test
since the $Ce^{4+}$ ion has no magnetic moment whereas the
$Ce^{3+}$ ion has 2.56$\mu_B$ magnetons"}. According to the
formula $\mu_{eff} = 2.84 \surd(\chi * T)$ they report the
following effective moments $\mu_{eff}$ for Ce metal at three
temperatures: at $90\,K: \mu_{eff} = 1.80\mu_B$, at $195\,K:
\mu_{eff} = 2.23\mu_B$ and at $291\,K: \mu_{eff} = 2.34\mu_B$.

\begin{figure}[tb]
\begin{center}
\includegraphics[width=18pc]{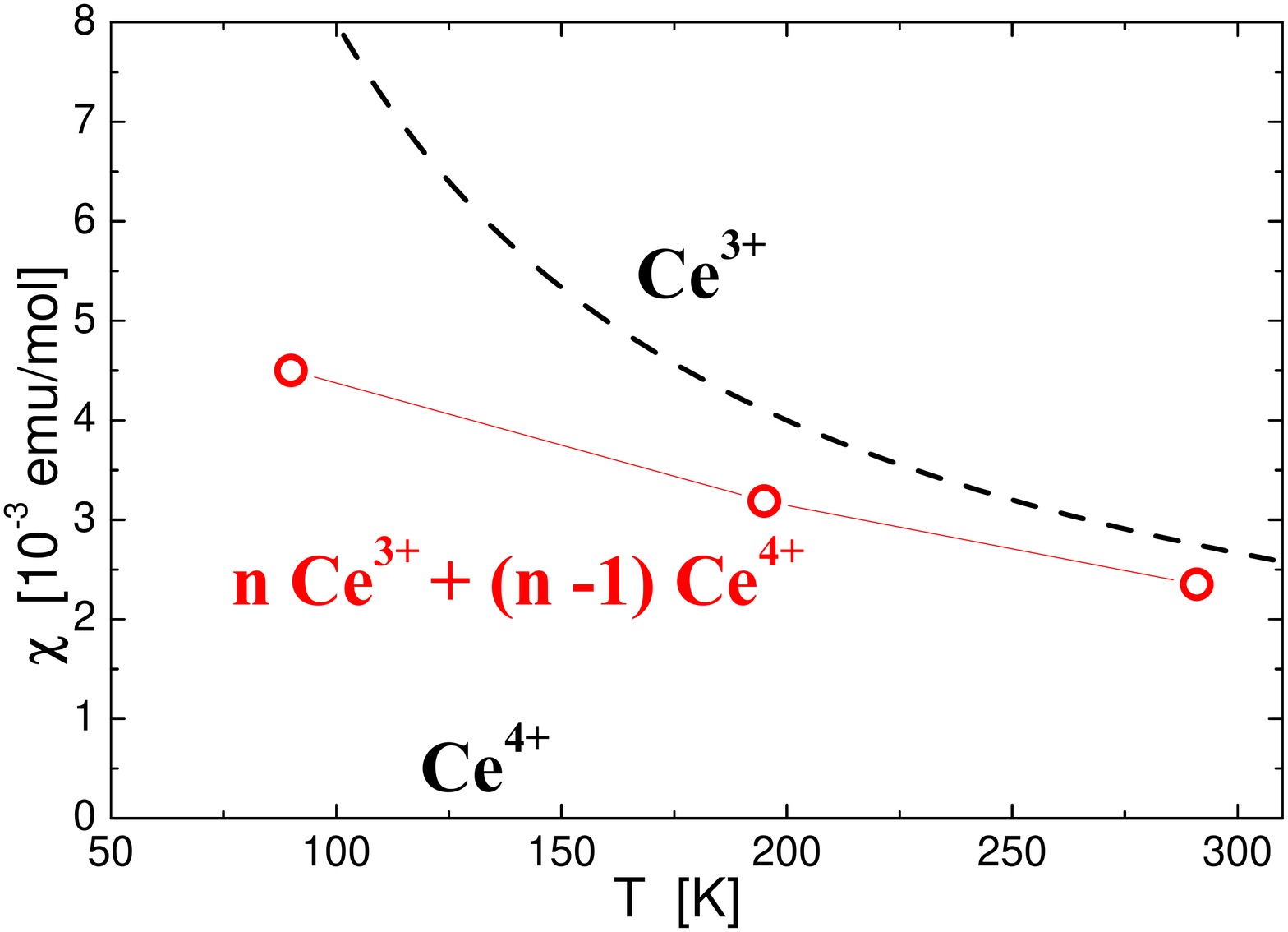}
\end{center}
\caption{Mean valence of Ce metal evaluated from the magnetic
susceptibility thermal dependence using a linear interpolation
(mixing rule) between Ce 3$^+$/4$^+$ type ions. Values after
\cite{Klemm37}.} \label{F10}
\end{figure}

After having extracted this temperature dependence of the
effective magnetic moments, they propose a description of their
results based in a mixing valence picture: {\it "... that's why
the actual moments lie between the two limit values. The values
closer to $Ce^{3+}$ are at higher the temperature. According to
the mixing rule (valid only as a first approach), from the
observed susceptibility values the ratio between $Ce^{4+}$ and
$Ce^{3+}$ ions is: i) both charges are similar at 90\,K; at 195\,K
$Ce^{3+}$ ions amounts to 70\%, and to 84\% at room temperature."}
as it is shown in Fig.~\ref{F10}.

\subsection{Cerium electronic configuration instability}

At the end of the following decade the early concepts on the
possible Cerium electronic instability were drown by different
authors. One of them in {\bf 1949} by A.W. Lawson and T-Y Tang
\cite{Lawson49} indicates that {\it "Powder patterns, taken at
approximately 15\,kbar, reveal that the high pressure modification
of the ambient pressure FCC lattice is also FCC!. The new
structure possesses a lattice constant $a = 4.84 \pm 0.03$\,\AA,
yielding an over-all volume change at this pressure of 16,5\%."}
These values remind those from Bridgman in 1948 \cite{Bridgman48}.

The microscopic origin of such a volume collapse is then proposed
{\it "as a result of stimulating conversations with our colleague,
W.H. Zachariasen, we propose the following simple model : .... Ce,
which is the first atom in the periodic table to permit the
existence of a $4f$ electron, when condensed exhibits the tendency
to become perverted from a 3-valent to a 4-valent state."}.

This breaking through idea refers to an electronic transition:
{\it "Apparently, the application of ~12\,kbar of pressure is
sufficient to evoke this transformation and the $4f$ electron is
literally squeezed into a $5d$ state."}.

Quite simultaneously (in {\bf 1950}) a similar electronic effect
proposed by Linus Pauling is quoted by A.F. Schuch and J.H.
Sturdivant \cite{Schuch50} indicating that in 1944 \cite{Trombe44}
{\it "... Trombe and Foex observed a transition at 109\,K
accompanied by a 10\% volume decrease and a decrease in magnetic
susceptibility. Prof. Linus Pauling  suggested to us in 1946 that
this transition is caused by the promotion of a $4f$ electron to a
bound-forming orbital."}. One may appreciate that in the late 40's
more quantum mechanic pictures emerged in the attempt to
understand the Ce-valence dilemma.

\subsection{Coqblin-Blandin model}

The well known Coqblin-Blandin model ({\bf 1968} \cite{Coqblin})
was a qualitative improvement to the knowledge of the microscopic
mechanism of Ce magnetic properties. An illustrative summary of
this model is provided in a chapter of the "Handbook of Phys. and
Chem. of Rare Earths" dedicated to "Cerium" \cite{Cerium}.

\begin{figure}[tb]
\begin{center}
\includegraphics[width=19pc]{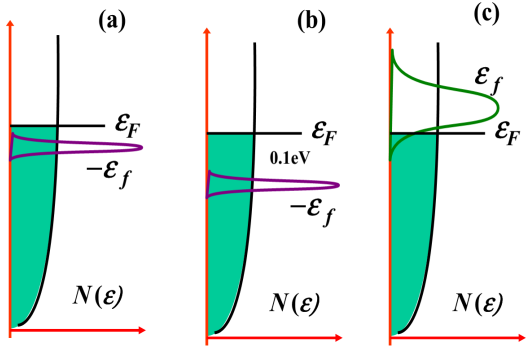}
\end{center}
\caption{a) Localized $4f$ state close in energy to the Fermi
level, b) $4f$ virtual bond state lying about 0.1\,ev below the
fermi level and having a half-width of around 0.02\,ev, c)
slightly occupied highly hybridized $4f$ level.} \label{F11}
\end{figure}

This model is described as {\it "One consequence of the
promotional idea is that it implies the localized $4f$ state in Ce
is close in energy to the Fermi level and therefore should also be
coincident with the conduction band}, see Fig.~\ref{F11}a. Their
feeling was that {\it "... under these conditions the $4f$ state
should mix or hybridize with the conduction band states to form a
$4f$ virtual bound state"}. This basic idea clearly reminds the
earlier proposition by Zachariasen-Pauling mentioned in Section
III-B.

Concerning the allotropic phases of Ce metal: {\it "In $\gamma-Ce$
the occurrence of localized moments could be described as arising
from a spin and orbital magnetic $4f$ virtual bond state lying
about 0.1\,ev below the fermi level and having half-width of
around 0.02\,ev}, see Fig.~\ref{F11}b. Then, {\it "The mechanism
for the transition to $\alpha-Ce$ was believed to arise from a
cooperative interaction between a number of volume dependent terms
in their free energy expression"}, as depicted in Fig.~\ref{F11}c.

\subsection{Phenomenological models}

Through measurements of magnetic, structural, transport and
spectroscopic properties on RE compounds it has been established
that nonmagnetic RE ions can be described as fluctuating in time
between two electronic configurations \cite{ICF}. In SmB$_6$, SmS,
YbAl$_3$, $\alpha$-Ce, TmTe, and EuCu$_2$Si$_2$, each
configuration is characterized by a different integral occupation
of the $4f$ shell. In this "Inter-Configurational Fluctuations"
(ICF) model, it is assumed that {\it "...there are two distinct
states of the rare earth atom and its environment, one with an
integral number '$n$' of electrons in the $4f$ shell, the other
with '$n - 1$' electrons in the $4f$ shell together with one
delocalized electron in the environment"}. The energies of these
two states are $E_n$ and $E_{n-1}$ respectively, and one defines
$E_{ex}=E_n-E_{n-1}$. Thus {\it "In each state $E_i$ the $4f$
shell is assumed to have a well defined angular momentum 'J' and a
corresponding magnetic moment '$\mu_i$'"}. Although this model
properly described the magnetic dependence of Yb, Eu and Sm
compounds \cite{ICF}, it is not applicable to Ce ones because the
energy difference between Ce$^{3+}$ and Ce$^{4+}$ electronic
configurations is about 60\,kCal/mol $>2$\,ev/mol \cite{Hformat}.


Other phenomenological approaches to the valence instability of
Cerium atoms were obtained analyzing the relationship between
crystal and structures in binary compounds. A clear correlation
was observed between cubic "fcc" structures and significant
hybridization of Ce-$4f$ states \cite{Sereni85}. In such symmetry
$p$-orbitals and $4f$ ones have the same parity and three of them
(among 7 $f$-orbitals) can be projected into the three formers. By
analyzing electronic configurations of the ligand band it was
found that the $'spdf'$ configuration may become the most
energetically favored after the metallization (or fusion) process.
For such computation the Gschneidner criterion \cite{GschBonding}
used to evaluate the contribution of each kind of electronic
orbital to the total cohesive energy was applied.

The most relevant example for the fcc-CeN. Its $p$-band character
was recognized though an unexpected contradiction between
$L_{III}$ and $L_I$ X-ray Absorption (XAS) edges measurements
\cite{Kappler91}. While the $L_{III}$ showed an extremely narrow
'white-line', the $L_I$ one displayed a double maximum in
agreement with the extreme mixed valence thermodynamic evidences
(specific heat, susceptibility, resistivity and thermal expansion
\cite{Olcese79}. Furthermore the Ce-volume collapse was revealed
by the $L_{III}$-XANNES themselves. The conclusion was that the
electron band has a 'pure-p' character and therefore the
($2\,p\to\epsilon\,d$) electronic transition, corresponding to the
$L_{III}$ absorption, is not allowed. On the contrary, the $L_{I}$
edge related to the ($2\,s\to\epsilon\,p$) transition occurs. This
restriction of the XAS spectroscopy explains most of the
contradictions between this technique and the thermodynamic
determination of the Ce valence in CeX compounds with X = a
semimetal of $p$-like character.

Another efficient mechanism to destabilize the valence of Ce-atoms
is provided by a chemical-potential 'pressure' which drives the
$4f^1$ electron transference from the Ce$^{3+}$ configuration to a
ligand-band hole like in CePd$_7$ \cite{CePd7}. In this compound,
the measured Sommerfeld coefficient: $\gamma(CePd_7) =
9.8$\,mJ/molK$^2$ equals the value of single $\gamma(Pd)$ atom.

Notably, the Ce-volume collapse and the related change of valence
($Z_i =$ 3 or 4) show a nearly constant relationship in the
product: $Z_i*Vol_i \approx 153$\,\AA$^3$, where $Vol_i$ is the
volume of the Ce-cell extracted from the respective 3$^+$ or 4$^+$
metallic radius \cite{Laves}.

\section{Conclusions}

The small spark of a Ceria stone that had lighted the darkness
along centuries already triggered the process of the Ce valence
transition. In the last century this valence instability was one
of the outstanding subjects in international conferences addressed
to the study of RE intermetallic compounds, that progressively
revealed the physical richness of this mechanism. Its relevance
was resumed in the statement by D.C. Koskenmaki and K. Gschneidner
Jr. \cite{Cerium}: {\it "In its elemental form Ce is the most
fascinating member of the Periodic Table"}.

\section*{Acknowledgments}
The author appreciate the Bernard Coqblin Prize awarded by the
{\it Strongly Correlated Electron Systems} (SCES) community for
the {\it "significant contribution to the physics of SCES achieved
in Argentina}, Prague, the 17th July of 2017.

\end{document}